\documentclass[]{aa}
\usepackage{graphicx}

\begin{document}

\title{A study of the stability regions in the planetary system HD 74156 -- Can it host earthlike planets in habitable zones?}

\titlerunning{Earthlike planets in HD 74155}

\author{R.\ Dvorak, E.\ Pilat-Lohinger, B.\ Funk and F.\ Freistetter}

\authorrunning{Dvorak et al.}

\offprints{R.\ Dvorak, \email dvorak@astro.univie.ac.at}

\institute{Institute for Astronomy, University of Vienna, 
      T\"urkenschanzstrasse 17, A-1180 Vienna, Austria}

\date{Received; accepted}   

\abstract{
Using numerical methods we thoroughly investigate the dynamical
stability in the region between the two planets found in HD 74156.
The two planets
 with  
semimajor axes 0.28 AU and 3.82 AU move on quite eccentric orbits
(e=0.649 and 0.354). There is a region between 0.7 and 1.4 AU which may host 
additional planets which we checked via 
numerical integrations using different dynamical models.
Besides the orbital evolution of several thousands of massless regarded planets
in a three-dimensional restricted 4-body problem 
(host star, two planets + massless bodies) we also
have undertaken test computation for the orbital evolution 
for fictitious planets with masses of 0.1, 0.3 and 1 $M_{\mathrm{JUP}}$
in the region between HD74156b and  HD74156c.
For direct numerical integrations up to $10^7$ years we used the Lie-integrator,
a method with adaptive step-size; additionally we used the Fast Lyapunov Indicators as tool
for detecting chaotic motion in this region. We emphasize the important r\^ole
of the inner resonances (with the outer planet) and the outer resonances (with
the inner planet) with test bodies located inside the resonances. In these two ``resonance'' regions
almost no orbits survive. The region between
the 1:5 outer resonance (0.8 AU) and the 5:1 inner resonance (1.3 AU), just in the right
position for habitability, is also not very likely to host planets.
Our results do not strictly ``forbid'' planets to move in the habitable zone, but their existence
is  rare.

}
\maketitle

\section{Introduction}
Since the discovery of the first extrasolar planetary system about 10 years ago a major point of dynamical investigations is the determination of
stable regions in extrasolar planetary systems where additional planets on stable orbits may exist. 
Today we have evidence for 
101 planetary systems with 116 planets, where 13 systems have more than one planet. The existence
of three (or more) massive bodies allows stability studies of the system itself; it is of special
interest when a planet is part of a double star system like in $\gamma$ Cephei
(Dvorak et al. 2003).
In this investigation we are interested in a system with 2 planets which could
permit other planets on stable orbits.
Therefore we have undertaken a dynamical study with the aid of numerical experiments of the 
G0 star HD 74156 hosting two planets on very eccentric orbits
(almost e=0.65 for the inner planet HD 74156b and e=0.4 for the outer planet HD 74156c)\footnote{we will use
in the following the suffix b for the inner and c for the outer planet}. 
As two recent publications show -- Barnes \& Raymond (2002)
speak of ``contain broad zones of stability'' whereas Menou \& Tabachnik (2003) found no surviving
planets in this region -- the problem of additional planets in HD74156 could not be answered satisfactory up to now.
Our investigation unveils the complicated structure of this region
containing stable {\em and} unstable
orbits close to each other and show that regular, chaotic and also sticky orbits exist in this domain.
 
In the following we discuss the dynamical models used and the method of establishing dynamical 
stability via long-term numerical integrations and chaos-indicators. Then we discuss the 
specific r{\^o}le of resonances in the region between the two planets from 0.44 AU to 2.4 AU 
caused by the inner planet (up to 0.8 AU) and the outer planet (down to 1.3 AU).
Next the domain between these resonances is studied
and the results of the stability studies for possible orbits of massless planets are shown. 
We also placed massive planets between 0.8 AU and 1.3 AU (also on inclined orbits) to check
their stability in this ``habitable zone'' (=HZ, Kasting 1993).

\section{The dynamical model and the methods}

The orbital parameters were taken from the Geneva group of observers\footnote{see http://obswww.unige.ch/$\sim$udry/planet/hd74156.html}          
with $a_\mathrm{b}=0.28$ AU, $e_\mathrm{b}=0.647$, the mass $M_\mathrm{b}=
1.61 M_\mathrm{Jupiter}$ and $a_\mathrm{c}=3.82$ AU, $e_\mathrm{c}=0.354$, $M_\mathrm{c}= 8.21 M_\mathrm{Jupiter}$.

Because of the large eccentricities of the two planets an analytical method to
solve this problem does not work;  any perturbation approach will fail. The method widely used by different
groups are numerical experiments in appropriate dynamical models. In our study we
used the following four different models:

\begin{itemize}
\item the restricted 4-body problem consisting of the star, the two planets and -- as test bodies --
massless planets in the same plane (={\bf Ia}) and on inclined orbits (={\bf Ib}).

\item the full 4 body-problem, where the additional planets had masses in the range between 
$0.1 M_{\mathrm{JUP}}$ to $1 M_{\mathrm{JUP}}$ on plane (={\bf IIa}) and on inclined
orbits (={\bf IIb})
\end{itemize}

The integration of the Newtonian equations of motion (the masses involved are point masses) 
was undertaken with the Lie-Integration method which
was already extensively used by our group (e.g. Hanslmeier \& Dvorak 1984;  
Lichtenegger 1984).
It uses an automatic step-size and is -- due to the recurrence of the Lie-terms -- a very fast
and precise integration method. 

As second independent tool 
to check the domains in between the two existing planetary orbits
we utilized the Fast Lyapunov Indicators (FLIs)
developed by Froeschl\'e et al. (1997). This program uses a
Bulirsch Stoer integration method and is especially adapted for distinguishing between chaotic and
regular orbits via the FLIs.

The integration times were in some experiments extended up to $10^8$ years. 
Our escape (stability) criterion was 
such that no close encounters within the Hill's sphere were allowed;
such an orbit was classified as
``unstable''. The criterion for the FLIs is somewhat different: the time
evolution of the FLIs defines clearly the orbital behaviour.
The two methods were used complementary and showed in most cases quite a good agreement 
of the results.

\section{The main resonance regions}

First of all we computed the domains in semimajor axes where the orbit of the inner (outer) planet
does not allow stable orbits, because the respective aphelion (perihelion) distance would lead to
close encounters. 
Thus the possible region for circular orbits without a crossing of the orbits 
shrinks to $0.44$ AU $\le  a \le 1.81$ AU.
It is well known from the main belt of asteroids that the mean motion
resonances are responsible for its structure with gaps and families. We therefore computed the
main resonances up to the 1:5 mean motion resonance with the inner planet 
(=outer mean motion resonance - OMMR) and
up to the 5:1 mean motion resonance with the outer planet (=inner mean motion
resonance - IMMR).
The main resonances which we have taken into account were the major ones known from
the asteroid belt:  2:1, 5:2, 7:3, 3:1, 4:1, 5:1 for the OMMR and the respective
resonances for the IMMR (1:5, etc.).
For motions in resonances the initial starting position is quite important, because sometimes
a protection mechanism avoids close approaches even for large eccentricities. We have checked
the IMMR and the OMMR carefully and have chosen 12 starting positions for the fictitious planet
in a certain resonance with the mean anomaly $M = 0^o, 30^o,...,360^o$.
The time of integration was $10^6$ years, but we emphasize that the extension from
$10^5$ years to $10^6$ years changed the results only slightly; therefore we are
convinced that the results are ``realistic''.

\begin{figure}
\centerline{
\includegraphics[height=8.15cm]{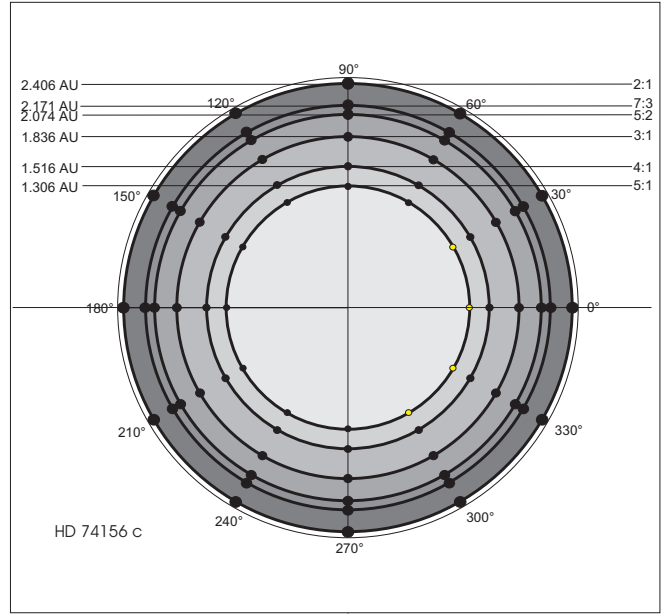}
}
\caption[]{Region of the IMMR in the dynamical model {\bf Ia} for different values
of the mean anomaly; the resonances with the outer planet (HD 74156c) are shown on the 
upper right side, the corresponding semimajor axes
on the upper left side. Filled circles mark unstable orbits and open circles mark stable orbits.}
\end{figure}

\begin{figure}
\centerline{
\includegraphics[height=7.5cm]{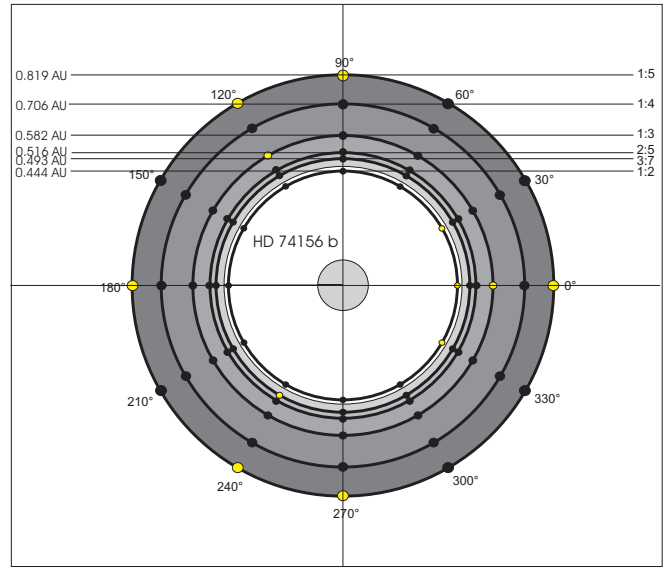}
}
\caption[]{Region of the OMMR in model {\bf Ia}; the respective resonances are with HD 74156b. 
Specifications like in Fig. 1} 
\end{figure}

\begin{figure}
\centerline{
\includegraphics[height=7.5cm]{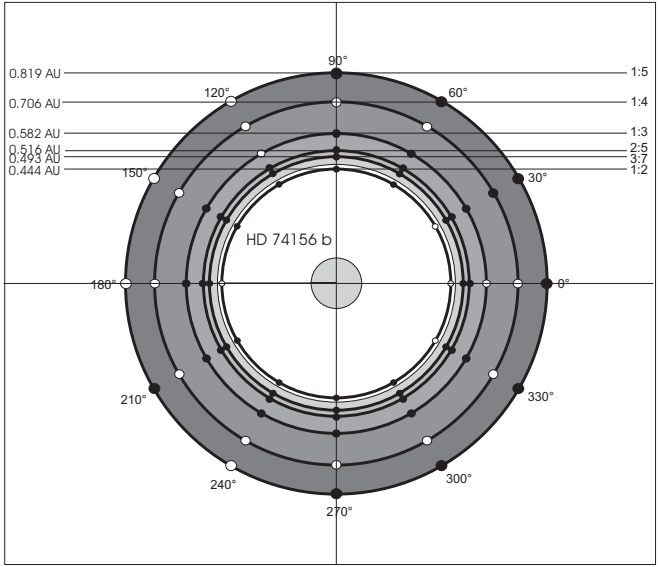}
}
\caption[]{Region of the OMMR in model {\bf Ib} ($i = 30^o$); the specifications are the same as in Fig. 1} 
\end{figure}

For the IMMR in model {\bf Ia} (Fig. 1) we can see that only a few orbits survive in the 5:1 resonance, 
all the others suffer from
close encounters with the outer planet sooner or later. 
We then studied also the IMMR in model {\bf Ib} for initial inclinations of the fictitious planet
between $10^o \le i \le 50^o$ with $\Delta i = 10^o$. Up to $i = 30^o$ the stability behavior did not
change significantly, for $i = 40^o$ out of the 72 orbits in the IMMR only $10\%$ survived 
in the higher resonances 4:1 and 5:1.

For the OMMR in model {\bf Ia} (Fig. 2) there are some orbits stable even in the 
1:2 resonance which overlap
-- concerning the distance to the host star --
with the aphelion distance of the inner planet. Their survival is due to the initial conditions chosen
for some orbits and is a well known fact.  The picture is different from
the one for the IMMR where only resonances far away from the planet's orbit survive; for the OMMR
we can see stable orbits in all resonances with the exception of the 2:5 and the 1:4 resonance.
In contrary to the IMMR the percentage of stable orbits in these resonances increases for
larger initial inclinations; only for $i = 50^o$ almost no orbit survives in this region.

As another check we started different fictitious massless planets in model {\bf Ia} and 
{\bf Ib} in between the resonances using the FLIs. As example we show the plot of an experiment where the 
test planet was started
with a=0.7 AU close the 1:3 resonance (Fig. 4a); the starting positions 
of the two planets were chosen between $0^o$ and $360^o$ (model {\bf Ia}). Only for a strip  
around $M_\mathrm{b} = 0$, where M is the mean anomaly, the orbits turned out to be stable; these results 
coincide with the stability of the
1:3 resonance of Fig. 2 for M = 0. As second example we show the dependence on the initial conditions for a
fictitious planet with an initial semimajor axis a = 1.3 AU (Fig. 4b), which is close to the 5:1 resonance.
A broad strip of stable orbits is visible for certain combinations of the initial conditions, about
33$\%$ of the orbits are unstable; here the agreement with Fig. 1 (the orbits of the 5:1 resonance
at a = 1.306 AU) is not so good; the ratio of stable orbits to unstable ones is almost vice-versa
(4 out of 12 are stable). This fact shows the complicated structure of the phase space, where
stable and unstable orbits are sometimes very close to each other according 
to high order resonances; they can be stable for quite long time on ``sticky orbits'' (e.g. Dvorak et al, 1998).
Globally one can say that stable orbits in this domain of IMMR and the inner part of the OMMR are 
quite rare.

\begin{figure}
\centerline{
\includegraphics[height=4.5cm]{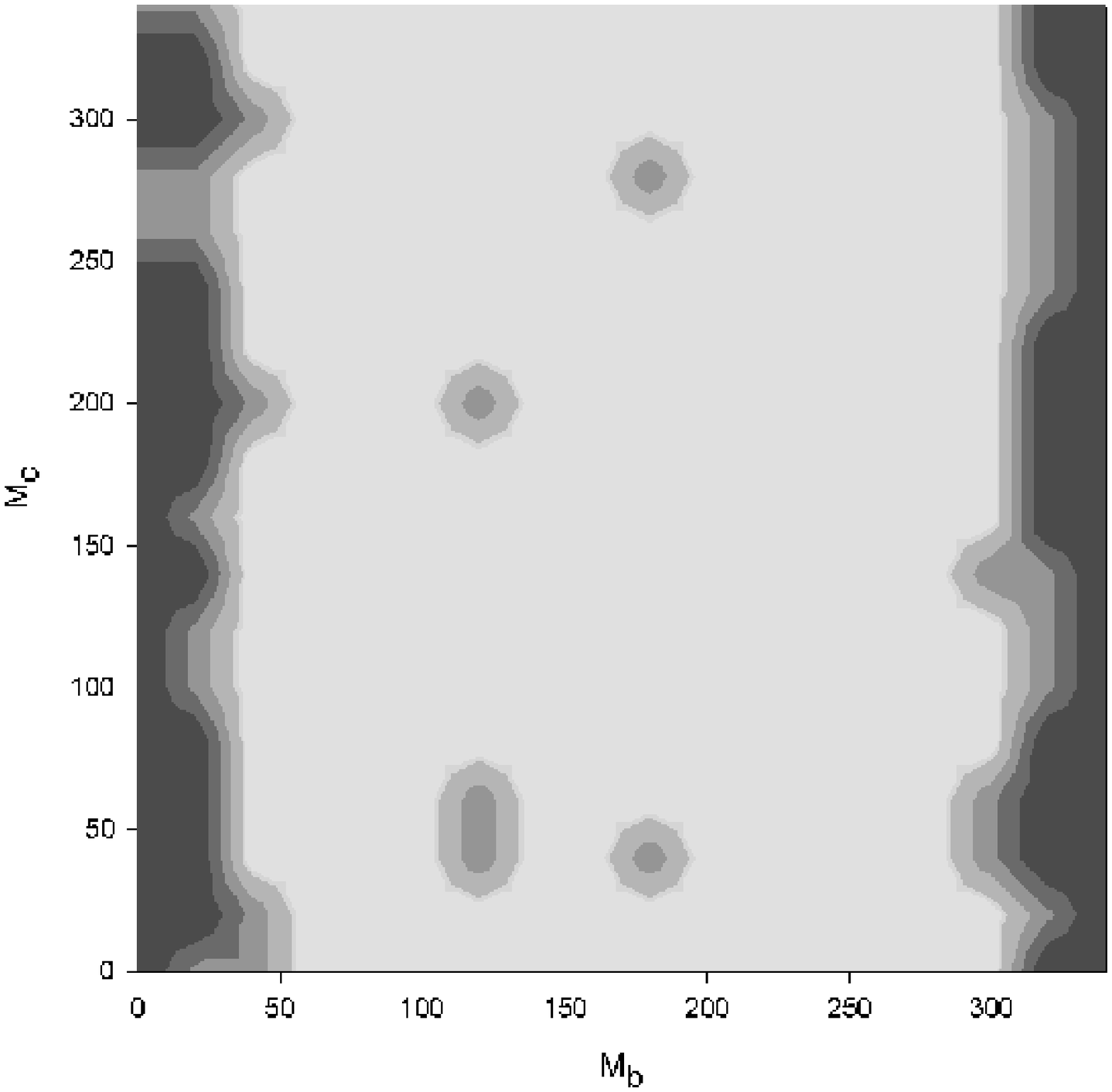}
\includegraphics[height=4.5cm]{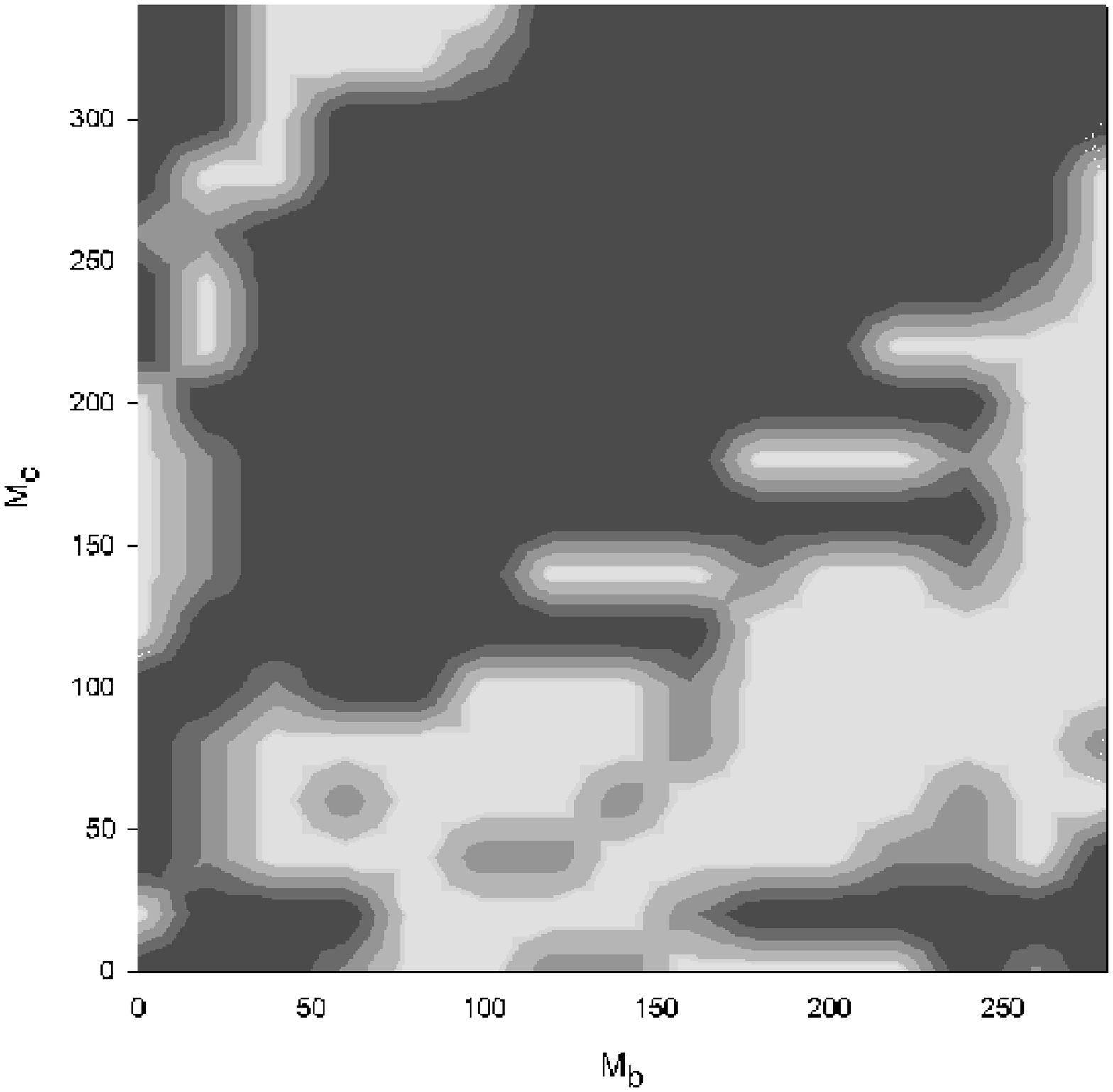}
}
\caption[]{Stability of a fictitious planet started in a circular orbit with a
  = 0.7 AU (left panel)
and a = 1.3 (right panel) for different
initial positions of the two perturbing planets; x-axis is the mean anomalie of the inner planet ($M_\mathrm{b}$),
y-axis is the mean anomalie of the outer planet ($M_\mathrm{c}$). Stable regions are marked as black regions according
to the values of the FLIs} 
\end{figure}

\section{The region between the resonances}

For the region in between 0.8 and 1.3 AU, the HZ, 
we wanted to check how the stability of the orbits there depends on the initial eccentricities of
the two planets (we did not change their semimajor axes, which seem to be quite well 
determined through the radial velocity curves). Therefore we have undertaken 
straightforward integrations in model {\bf Ia}
for four fixed values $e_{\mathrm{c}} = 0.3, 0.35, 0.4$ and $0.45$. For each diagram one parameter 
was the initial semimajor axis of the fictitious planet 
$0.6$ AU $\le a \le 1.4$ AU, the other one
was the eccentricity of the inner planet $0.5 \le e_{\mathrm{b}} \le 0.7$. The
actual orbital parameters
are quite well inside these ranges of eccentricities of the orbits of the two 
planets. Our criterion for the ``grade of stability'' was the eccentricity 
of the orbit of the fictitious planet staying within a certain range during the whole integration
 characterized with different grey tones (from dark to light grey), white
stands for escaping orbits.
The criterion for the darkest region with $e<0.2$ guarantees that the differences in distance to
the host star and thus the temperature changes are relatively small and favor
the development and stability of atmospheres (Lammer, 2003, private communication)

\begin{figure}
\centerline{
\includegraphics[height=4.5cm]{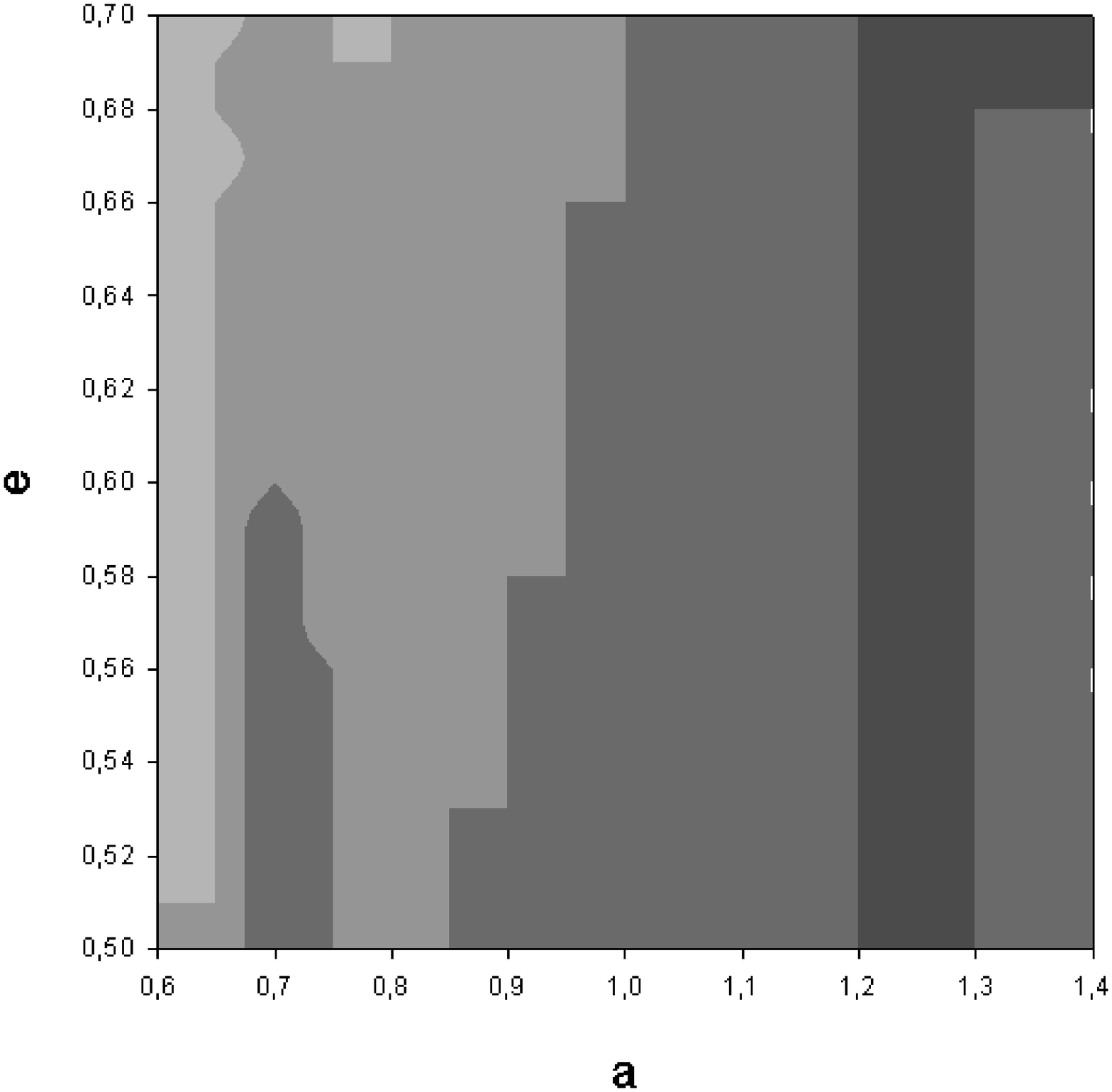}
\includegraphics[height=4.5cm]{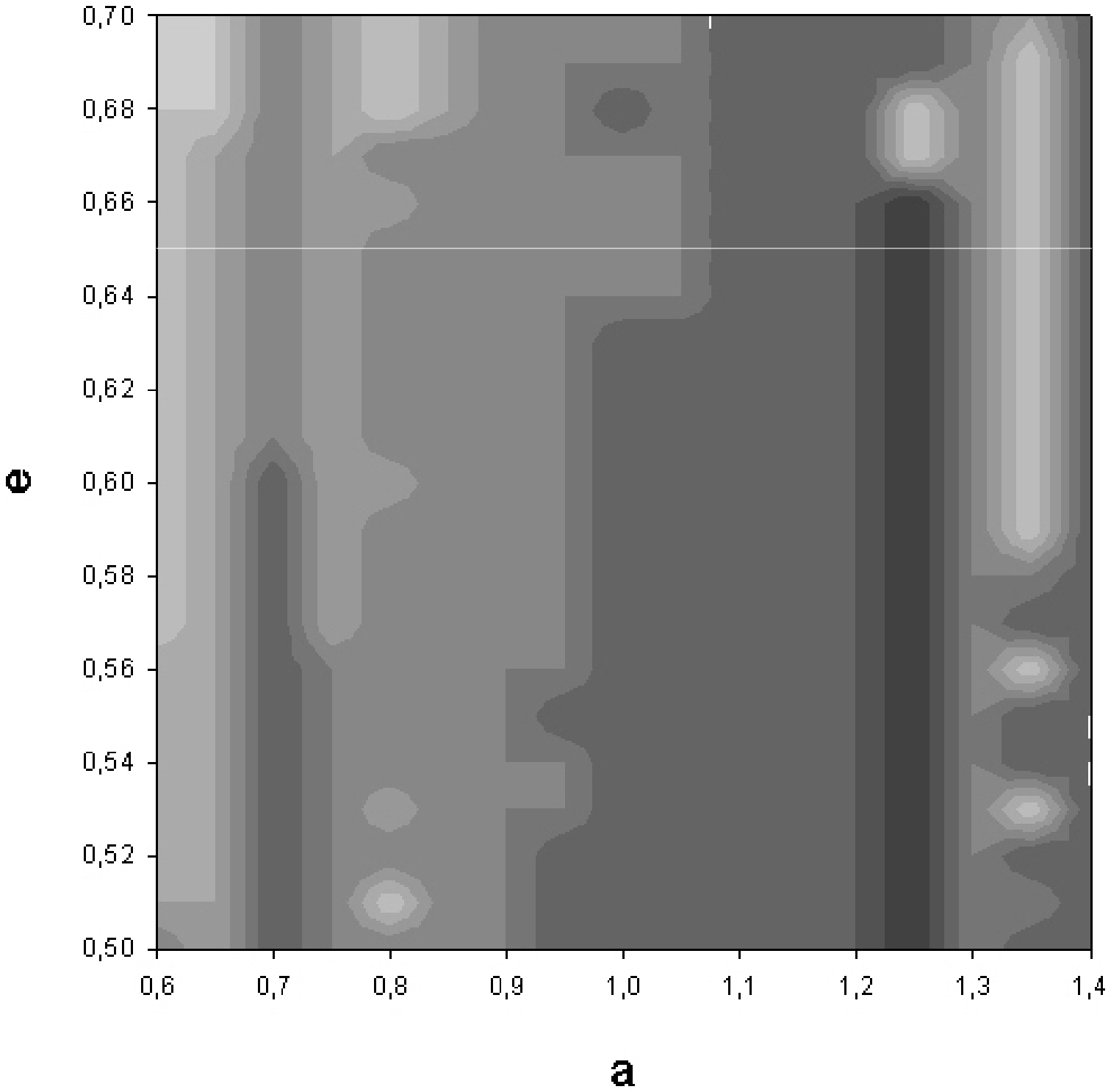}
}
\centerline{
\includegraphics[height=4.5cm]{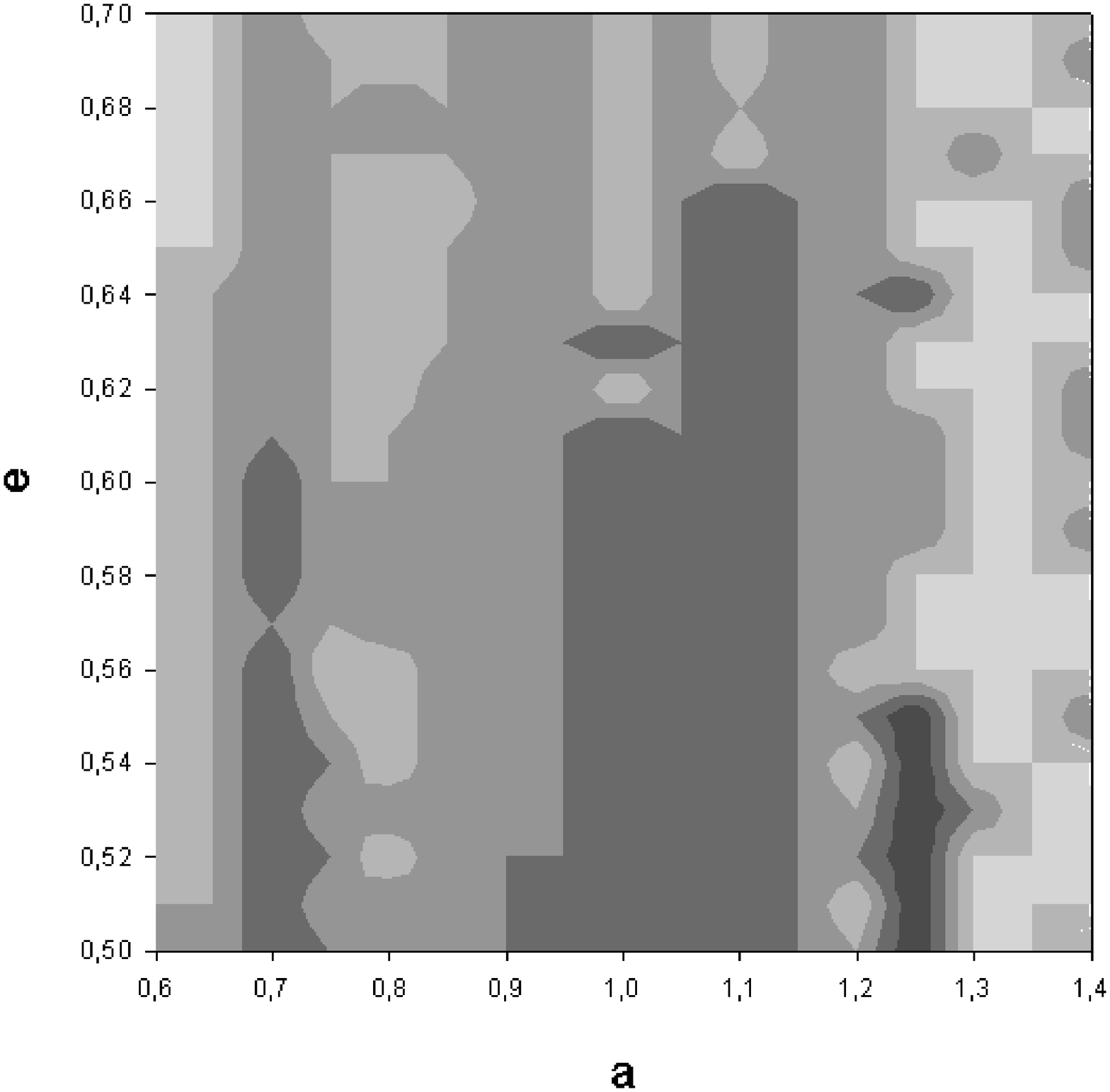}
\includegraphics[height=4.5cm]{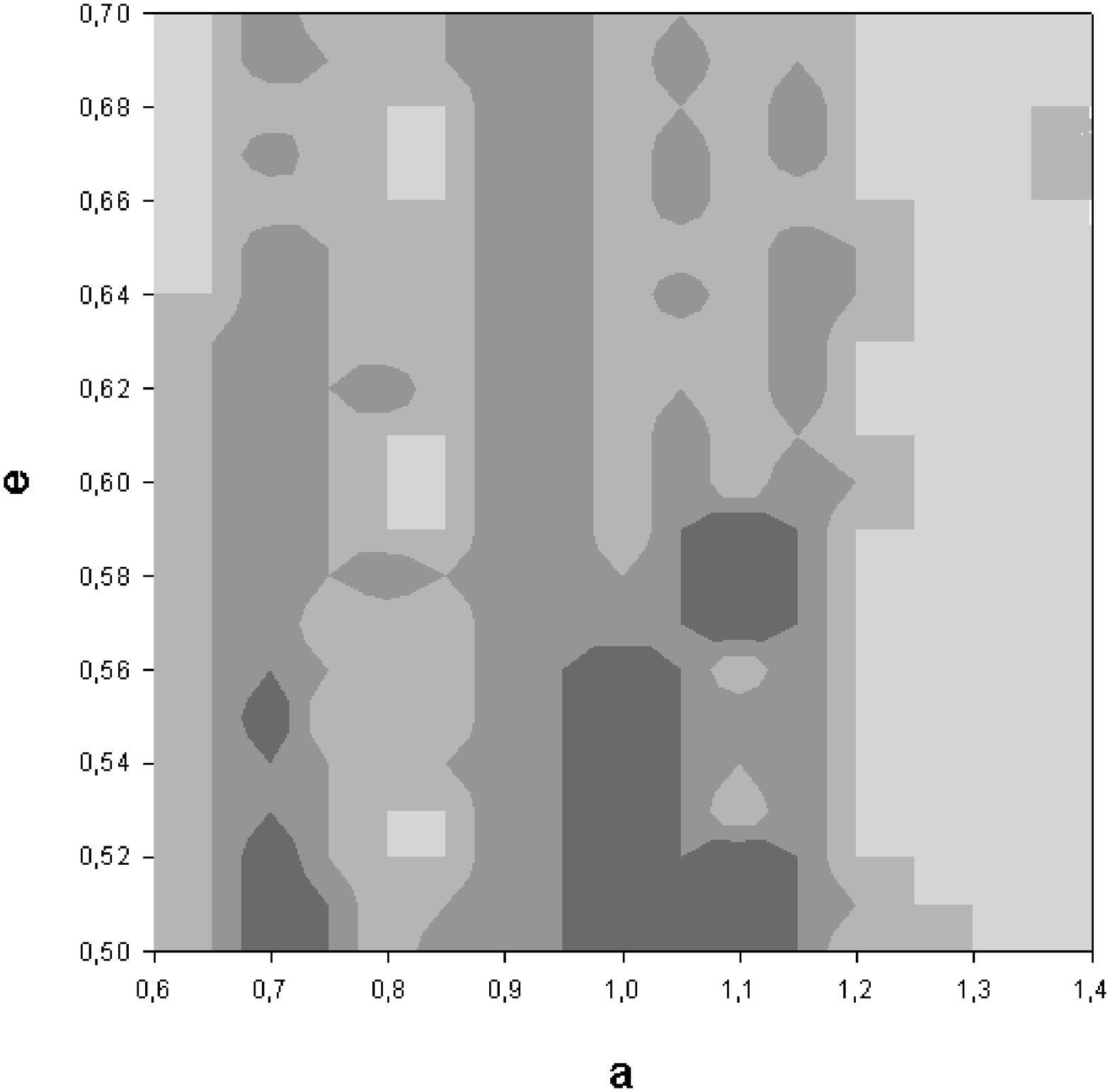}
}
\caption[]{Regions of motion for a fictitious planet between the inner planet and the outer planet
of HD 74156 for an eccentricity $e_{\mathrm{c}} = 0.3$ (upper left graph),
$e_{\mathrm{c}} = 0.35$ (upper right graph), $e_{\mathrm{c}} = 0.4$ (lower left graph) and
 $e_{\mathrm{c}} = 0.45$ (lower right graph)
for the outer planet. Abscissa is the eccentricity of the
inner planet (with an actual value of 0.649), ordinate is the initial semimajor axes of the 
fictitious planet always started in a
circular orbit, dark regions are stable; grey stands for the maximum eccentricity during the
whole integration time: black means $e < 0.2$, white stands for escaping orbits}
\end{figure}

For $e_{\mathrm{c}} = 0.3$ (Fig. 5a) -- besides the ``finger'' around $a=0.7$ AU (1:4 resonance) --  
we see a broad stability region between 0.9 and 1.4 AU where the eccentricities are $e<0.3$; it becomes slightly 
smaller for larger $e_\mathrm{b}$. At $a=1.25$ AU, close to the 16:3 resonance, there exists a stable strip of orbits 
with small eccentricity variations ($e<0.15$). This strip is also visible in Fig. 5b ($e_\mathrm{c}=0.35$),   
where $e_\mathrm{b} = 0.649$ is marked by a straight line. In the next 2 figures (Fig. 5c for $e_\mathrm{c}=0.4$ and
Fig. 5d for $e_\mathrm{c}=0.45$) we see how the stability of the orbits weakens with increasing eccentricity of the outer planet. 
Whereas in Fig. 5c we still see a small finger of very stable orbits at $a=1.25$ AU, in Fig. 5d no orbits 
of the fictitious planets for $a \le 1.2$ AU survive!  

We also checked how secular resonances may act in this region between OMMR and IMMR\footnote{Secular resonances
arise when the motion of the perihelion (Keplerian element $\omega$) or the motion of the
longitude of the node (Keplerian element $\tilde \omega$) is in resonance with the respective motion of the
regarded planet.}. It turned out that both planets have periods in the order of some
$10^4$ years while for planets in the HZ these frequencies are in the order of only some $10^3$ years. We conjecture that 
the reason for chaotic motions between the planets b and c are 
the recently discovered three-body-resonances (e.g. Morbidelli 2002) which are important for
asteroids in the main belt -- but this will be investigated in more detail in a future study.

\section{Massive planets}

Further computations have been undertaken in model {\bf IIa} and {\bf IIb}
for the region between IMMR and OMMR for three different masses of the additional planet, namely 0.1, 0.3 and
1 $M_{\mathrm{JUP}}$. 
The orbital character of these planets was not very different from the ones of the non-massive 
fictitious planets:
only very few orbits were found to be stable (especially around a = 1.1 AU), but most of them were unstable for larger
inclinations. 
 This is somewhat in contradiction to the paper already cited
above (Barnes \& Raymond 2002), where the authors claim that they have found  ``broad zones of stability''. Unfortunately
it is not possible to compare the results directly because up to now only an abstract paper is available.
We plan a complete study, where massive planets are involved, using the
Lie-code in a parallel version (Stadel et al, 2003).

\section{Conclusions}

In this study we checked how the motion in resonances destroys certain stability regions 
outside planet b (OMMR) and inside planet c in the extrasolar system HD 74156. Furthermore the small
region between 0.8 AU and 1.3 AU was studied through  numerical experiments using the orbits of thousands of 
fictitious planets there. 
The main result is that there can survive orbits for long time intervals, but they seem to be quite rare and
depend on very special initial conditions. We found orbits surviving for more
than $10^7$ years for different initial 
semimajor axes, but very close to these orbits others do not survive $10^5$ years. This typical
behavior of stickiness is well known in regions were regular and chaotic motions are mixed. 
For initial conditions close to a=1.25 the orbits stay within the eccentricity range $e \le 0.12$, thus well inside the HZ
of HD 74156.

\begin{acknowledgements}
E.\ Pilat-Lohinger wishes to acknowledge the support by the Austrian FWF
(Hertha Firnberg Project T122). B.\ Funk and F.\ Freistetter wish to 
acknowledge the support by the Austrian FWF (Project P14375-TPH).
Special thanks go to Markus Gyergyovits, who draw our attention to
this problem and Ruth Gr{\"u}tzbauch, who helped editing this article.
\end{acknowledgements}

\end{document}